\documentclass[twocolumn,showpacs,preprintnumbers]{revtex4}
\usepackage{amsmath}
\usepackage[dvips]{graphicx}
\usepackage{bm}

\begin{document}

\title{Magnetic phase diagrams of barcode-type nanostructures}
\author{B. Leighton$^{1,2}$}
\author{O. J. Suarez$^3$}
\author{P. Landeros$^3$}
\author{J Escrig$^{1,2}$}
\affiliation{$^1$ Departamento de F\'{\i}sica, Universidad de Santiago de
Chile (USACH), Avenida Ecuador 3493, 917-0124 Santiago, Chile\\
$^2$Centro para el Desarrollo de la Nanociencia y Nanotecnolog\'{\i}a,
CEDENNA, 917-0124 Santiago, Chile\\
$^3$ Departamento de F\'{\i}sica, Universidad T\'{e}cnica Federico Santa Mar%
\'{\i}a, Avenida Espa\~{n}a 1680, Casilla 110 V, 2340000 Valpara\'{\i}so,
Chile}

\begin{abstract}
The magnetic configurations of barcode-type magnetic nanostructures
consisting of alternate ferromagnetic and nonmagnetic layers arranged within
a multilayer nanotube structure are investigated as a function of their
geometry. Based on a continuum approach we have obtained analytical
expressions for the energy which lead us to obtain phase diagrams giving the
relative stability of characteristic internal magnetic configurations of the
barcode-type nanostructures.
\end{abstract}

\maketitle

\section{Introduction}

Magnetic nanoparticles have attracted increasing interest among researchers
of various fields due to their promising applications in hard disk drives,
magnetic random access memory, and other spintronic devices. \cite%
{SMW+00,KDA+98,CKA+99,WAB+01,GBH+02} In addition, these magnetic
nanoparticles can be used for potential biomedical applications, such as
magnetic resonance imaging (the nanoparticles can be used to trace
bioanalytes in the body), cell and DNA separation, and drug delivery. \cite%
{ET03} To apply nanoparticles in various potential devices and
architectures, it is very important to control their size and shape in order
to keep the thermal and chemical stability. \cite{PKA01}

The trusty sphere remains the preferred shape for nanoparticles but this
geometry leaves only one surface for modification, complicating the
generation of multifunctional particles. Thus, a technology that could
modify differentially the inner and outer surfaces would be highly
desirable. \cite{Eisenstein05} Tubular nanostructures have stimulated
extensive research efforts in recent years because of their particular
significance for prospective applications. A wide range of materials
including semiconductors, polymers, and metals have been prepared in the
form of nanotubes. \cite{YSN+04,WZC+04,MYW+04,BJK+07,EBJ+08} Although the
magnetic nanotubes has been intensely investigated, barcode-type
nanostructures have received less attention, in spite of tailoring the
multisegmented nanotube structure, along with the functionalization of the
inner wall surface of barcode-type nanotubes with various molecules (for
example, proteins and DNA). Moreover, they are expected to be particularly
useful in the field of catalysis, advanced microfluidics, molecule
separation and biological and magnetic sensors as well. \cite%
{LSN+05,NGR+01,Lehmann02,NCS+03,SCM+06,SRH+05} It is worth to mention that
barcode-type magnetic nanostructures consisting of regular arrays of
magnetic segments have been considered as providing the basis for extending
magnetic storage densities beyond the superparamagnetic limit. In such
system, a single tube with $n$ magnetic layers might store up to 2$^{n}$
bits, whose volume is much larger than those of the grains in conventional
recording media, bearing this way thermal fluctuations and increasing the
recording density by a factor 2$^{n-1}$. \cite{AHM+05,ELA+06}\ Recently, 
\cite{LSN+05} the preparation of metallic nanotubes based on the
preferential electrodeposition of a metal along the pore walls of an anodic
alumina oxide (AAO) membrane, in the presence of metallic nanoparticles on
the wall surfaces, has been reported. In the paper by Lee\textit{\ et al}. 
\cite{LSN+05} they were able to prepare multisegmented metallic nanotubes
with a bimetallic stacking configuration along the tube axis, showing
different magnetic behavior as compared with continuous ones, which
encourage a study about the possible magnetic configurations and
magnetostatic interactions in these barcode-type magnetic nanotubes.
Clearly, for the development of magnetic devices based on those arrays,
knowledge of the internal magnetic structure of the barcode-type
nanostructures is of fundamental importance.

The purpose of this paper is to investigate the magnetic ordering of
barcode-type nanostructures. Our particles are characterized by a set of
geometrical parameters, as depicted in Fig. 1. First, their external and
internal radii, $R$ and $a$, respectively, and total length, $L$, which
includes the magnetic matter as well as the nonmagnetic one. It is
convenient to define the ratio $\beta \equiv a/R$, so that $\beta =0$
represents a solid cylinder and $\beta \rightarrow 1$ corresponds to a very
narrow tube. We denote with $W$ the length of each ferromagnetic segment and
with $d$ the length of each nonmagnetic portion, so that, if we have a
barcode-type structure with $n$ magnetic tubes, the total length can be
written as $L=nW+\left( n+1\right) d$. 
\begin{figure}[h]
\begin{center}
\includegraphics[width=8cm]{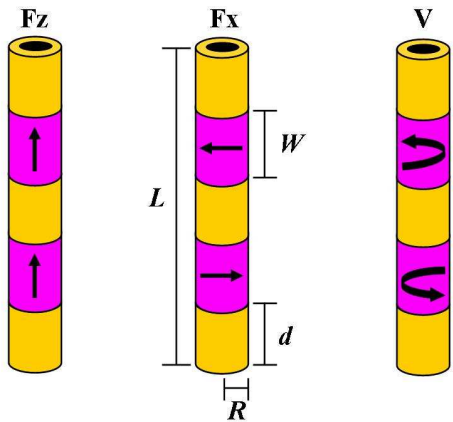}
\end{center}
\caption{Geometrical parameters and magnetic configurations of barcode-type
nanostructures.}
\end{figure}

\section{Model and discussion}

We adopt a simplified description of the magnetic system in which the
discrete distribution of magnetic moments is replaced by a continuous one,
defined by the magnetization vector field $\mathbf{M}(\mathbf{r})$ such that 
$\mathbf{M}(\mathbf{r})\delta v$ gives the total magnetic moment within an
elementary volume $\delta v$ centered at $\mathbf{r}$. The total magnetic
energy ($E_{tot}$) is generally given by the sum of four terms: exchange,
dipolar, anisotropy and Zeeman contributions, which are taken from the well
known continuum theory of ferromagnetism. \cite{Aharoni96} As we are
interested in the study of the relative stability of the zero-field magnetic
ground states, the contribution of the Zeeman energy can be disregarded.
Under these assumptions, the magnetic energy is just given by the dipolar ($%
E_{dip}$), exchange ($E_{ex}$), and anisotropy ($E_{k}$) contributions.

The total magnetization can be written as $\mathbf{M}\left( \mathbf{r}%
\right) =\sum_{i=1}^{n}\mathbf{M}_{i}\left( \mathbf{r}\right) $, where $%
\mathbf{M}_{i}\left( \mathbf{r}\right) $ is the magnetization of the $i$-th
ferromagnetic segment. In this case, the magnetostatic potential $U\left( 
\mathbf{r}\right) $ splits up into $n$ components, $U_{i}\left( \mathbf{r}%
\right) $, associated with the magnetization of each ferromagnetic segment.
Then, the total dipolar energy can be written as $E_{dip}=\sum_{i=1}^{n}$ $%
E_{dip}\left( i\right) +\sum_{i=1}^{n-1}\sum_{j=i+1}^{n}E_{int}\left(
i,j\right) $, where 
\begin{equation*}
E_{dip}\left( i\right) =\frac{\mu _{0}}{2}\int \mathbf{M}_{i}\left( \mathbf{r%
}\right) \cdot \nabla U_{i}\left( \mathbf{r}\right) dv ,
\end{equation*}%
is the dipolar contribution to the self-energy of the $i$-th ferromagnetic
segment, and 
\begin{equation*}
E_{int}\left( i,j\right) =\mu _{0}\int \mathbf{M}_{i}\left( \mathbf{r}%
\right) \cdot \nabla U_{j}\left( \mathbf{r}\right) dv ,
\end{equation*}%
is the dipolar interaction between ferromagnetic segments $i$ and $j$.

Usually, the exchange energy $E_{ex}$ in multilayer nanostructures has
contributions from both, the direct exchange interaction within the magnetic
segments and the other from the indirect interaction between them mediated
by the conduction electrons in the nonmagnetic layers. Since the indirect
interaction decays rapidly with the thickness of the nonmagnetic segment, it
can be neglected provided $d$ is large enough. A good estimate of the range
of the indirect exchange interaction can be obtained from the results for
multilayers. \cite{BJV+94} As a general result, the interlayer exchange
coupling vanishes for spacer thicknesses greater than a few nanometers,
which does not exceed the value of the exchange length $l_{x}=\sqrt{2A/\mu
_{0}M_{0}^{2}}$ of ferromagnetic metals. Here we focus our attention on
those cases in which $d$ is not smaller than the magnetic material's $l_{x}$%
, as the tubes fabricated by Lee \textit{et al}. \cite{LSN+05} which satisfy 
$d$ $\gg $ $l_{x}$ and thus interlayer exchange coupling can be safety
neglected. Therefore, to a good approximation can be written as $%
E_{ex}=\sum_{i=1}^{n}$ $E_{ex}\left( i\right) $, where $E_{ex}\left(
i\right) =A\int \left[ \left( \nabla m_{ix}\right) ^{2}+\left( \nabla
m_{iy}\right) ^{2}+\left( \nabla m_{iz}\right) ^{2}\right] dv$. Here, $%
\mathbf{m}_{i}=\left( m_{ix},m_{iy},m_{iz}\right) =\mathbf{M}_{i}/M_{0}$ is
the magnetization normalized to the saturation magnetization $M_{0}$ and $A$
is the stiffness constant of the magnetic material.

The cubic anisotropy energy of the particle can be added by means of the
following expression:%
\begin{equation*}
E_{c}\left( i\right) =K_{c}\int \left(
m_{ix}^{2}m_{iy}^{2}+m_{iy}^{2}m_{iz}^{2}+m_{iz}^{2}m_{ix}^{2}\right) dv,
\end{equation*}%
and the uniaxial anisotropy energy is given by%
\begin{equation*}
E_{u}\left( i\right) =-K_{u}\int m_{iz}^{2}dv.
\end{equation*}

On the basis of the above results, the total energy of the barcode-type
nanostructure can be written as $E_{tot}=\sum_{i=1}^{n}E_{self}(i)+%
\sum_{i=1}^{n-1}\sum_{j=i+1}^{n}E_{int}\left( i,j\right) $, where $%
E_{self}(i)=$ $E_{dip}(i)+E_{ex}(i)+E_{k}\left( i\right) $ is the
self-energy of the ferromagnetic segment $i$, and $E_{int}$ is the (dipolar)
interaction energy between two magnetic segments. We will proceed to
describe the magnetization of the different states we are considering here
and then we will evaluate the magnetic energy of each configuration. Results
will be given in units of $\mu _{0}M_{0}^{2}l_{x}^{3}$, i.e., $\tilde{E}%
=E/\mu _{0}M_{0}^{2}l_{x}^{3}$.

\subsection{Magnetic configurations}

It has been shown recently that single magnetic nanorings present three
basic ground states depending on their geometry (see Fig. 1). \cite{BLS+06,
LEA+06} These configurations are: (Fz) a quasi uniform magnetization state
oriented in the direction parallel to the cylindrical axis ($z$ axis); (Fx)
a quasi uniform magnetization state oriented in the plane perpendicular to
the $z$ axis; and (V) a flux-closure vortex state. For long nanorings $%
\left( W\gg R\right) $, the Fx phase is not present, \cite{BLS+06, LEA+06} a
result that holds for nanotubes. \cite{ELA+07, ELA+072}

It has been shown for Rothman \textit{et al}. \cite{RKL+01} that for
magnetized nanorings the single-domain in-plane ground state is the \textit{%
onion} state. Besides, the single-domain axial state for rings with small
inner diameter might be similar to the \textit{flower} state expected in
thick axially magnetized cylinders. \cite{UP04} Therefore, it may appear
questionable to select the uniform in-plane and axial as single-domain
states to build the phase diagram upon, as they are not stable
configurations at a zero applied field. However, it has been verified by micromagnetic
simulations \cite{BLS+06} and analytical calculations \cite{LEA+06} that the
energy difference between the actual single-domain ground state in a
nanoring, and the uniform state, often turns out to be very small. From this
available evidence, we conclude that the replacement of the more correct
quasi-uniform states by simpler ideal uniform states, only results in
uncertainties on the exact location of the phase boundaries and on some
physical values extracted from the phase diagram.

\subsubsection{Fz state}

For the Fz state, where the magnetization of the $n$ ferromagnetic segments
is uniform and parallel to the $z$ axis, $\mathbf{M}\left( \mathbf{r}\right) 
$ can be approximated by $M_{0}\hat{z}$, where $\hat{z}$ is the unit vector
parallel to the axis of the nanotube. In this case the exchange contribution
to the self energy vanishes, and the reduced self energy takes the form \cite%
{ELA+07}\ 
\begin{multline*}
\tilde{E}_{self}^{Fz}=\frac{\pi R^{3}}{l_{x}^{3}}\int_{0}^{\infty }\frac{dq}{%
q^{2}}\left( 1-e^{-q\frac{W}{R}}\right) \left( J_{1}\left( q\right) -\beta
J_{1}\left( q\beta \right) \right) ^{2} \\
-\frac{\kappa _{u}}{2}\frac{\pi WR^{2}}{l_{x}^{3}}\left( 1-\beta ^{2}\right)
,
\end{multline*}%
where $J_{1}\left( z\right) $\ is a Bessel function of the first kind and $%
\kappa _{u}=2K_{u}/\mu _{0}M_{0}^{2}$. In order to calculate the interaction
energy between the ferromagnetic segments, we first need to calculate the
magnetostatic potential $U\left( \mathbf{r}\right) $ of a single tubular
structure. The expression for this potential has been previously reported 
\cite{EAA+08} and is given by%
\begin{multline*}
U\left( r,z\right) =\frac{M_{0}}{2}\int_{0}^{\infty }\frac{dk}{k}J_{0}\left(
kr\right) \left[ RJ_{1}\left( kR\right) -aJ_{1}\left( ka\right) \right]  \\
\left( e^{-k\left\vert \frac{W}{2}-z\right\vert }-e^{-k\left\vert \frac{W}{2}%
+z\right\vert }\right) .
\end{multline*}%
From this equation it is possible to obtain the expression for the
magnetostatic field. Thus we write, $\mathbf{H}\left( r,z\right) =-\mathbf{%
\nabla }U\left( r,z\right) =H_{r}\left( r,z\right) \hat{r}+H_{z}\left(
r,z\right) \hat{z}$ with%
\begin{multline*}
H_{r}\left( r,z\right) =\frac{M_{0}}{2}\int_{0}^{\infty }dkJ_{1}\left(
kr\right) \left[ RJ_{1}\left( kR\right) -aJ_{1}\left( ka\right) \right]  \\
\left( -e^{-k\left\vert \frac{W}{2}-z\right\vert }+e^{-k\left\vert -\frac{W}{%
2}-z\right\vert }\right) 
\end{multline*}%
and%
\begin{equation*}
H_{z}\left( r,z\right) =\frac{M_{0}}{2}\int_{0}^{\infty }dkJ_{0}\left(
kr\right) \left[ RJ_{1}\left( kR\right) -aJ_{1}\left( ka\right) \right]
Y\left( W,z\right) ,
\end{equation*}%
where $Y\left( W,z\right) =sign\left( \frac{W}{2}-z\right) e^{-k\left\vert 
\frac{W}{2}-z\right\vert }-sign\left( -\frac{W}{2}-z\right) e^{-k\left\vert -%
\frac{W}{2}-z\right\vert }$. The function $sign\left( x\right) $ gives $-1$, 
$0$ or $1$ depending on whether $x$ is negative, zero, or positive. Figure 2
illustrates the magnetostatic field profile calculated analytically for
nanotubes with the same geometrical parameters as the ones investigated
experimentally by Lee \textit{et al.} \cite{LSN+05}. 
\begin{figure}[h]
\begin{center}
\includegraphics[width=8cm]{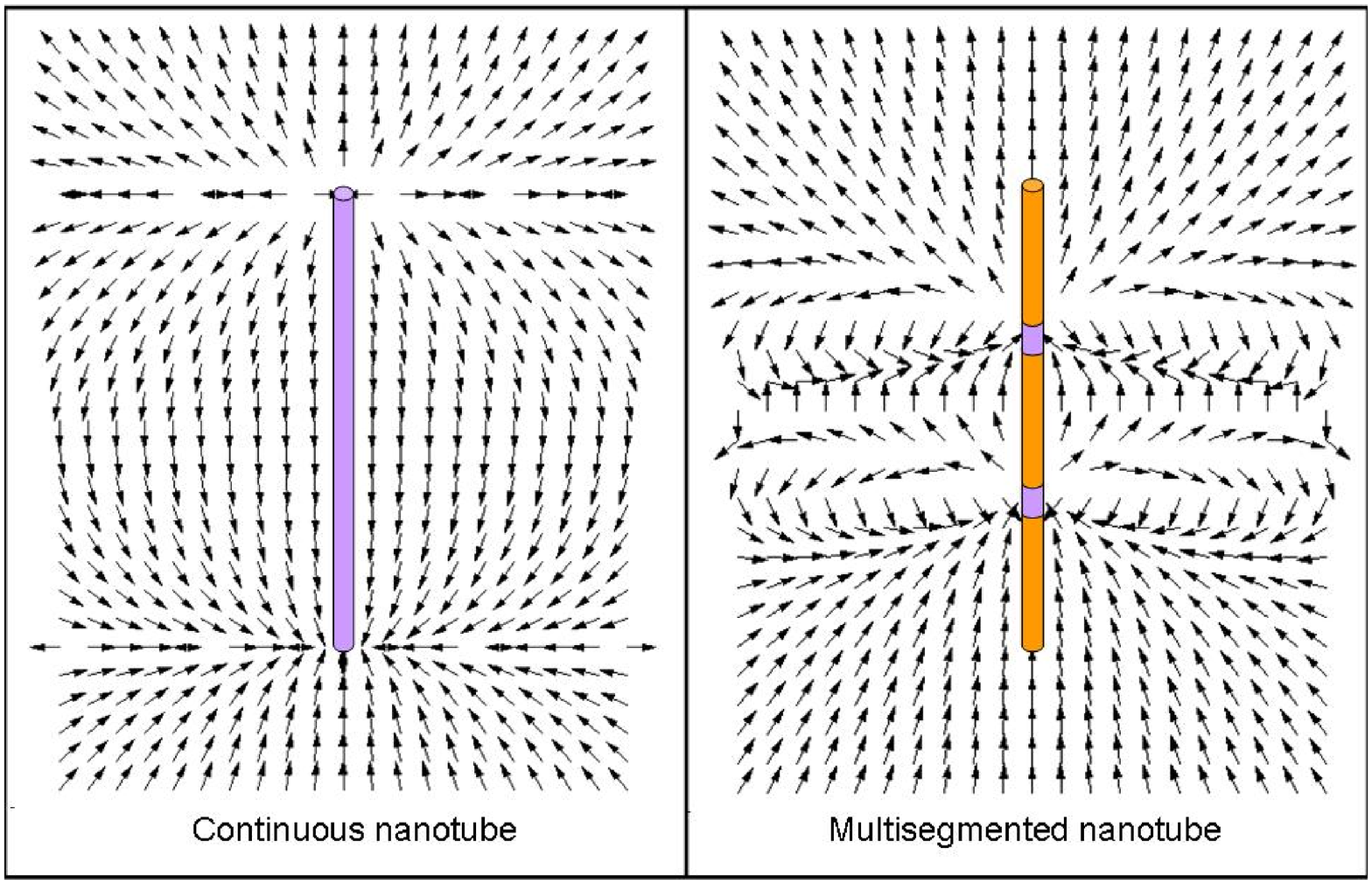}
\end{center}
\caption{Stray field direction (arrows) generated by a single nanotube 
\textit{(left picture)} and a multisegmented nanotube \textit{(right picture)%
} magnetized in the $+z$ direction. These two examples correspond to
nanotubes with the same geometrical parameters as those studied
experimentally by Lee \textit{et al }\protect\cite{LSN+05}. }
\end{figure}
Finally, the reduced interaction energy between two tubular nanostructures
has been calculated in rather general way by Escrig \textit{et al.} \cite%
{ELA+06,EAA+08} and is given by%
\begin{multline*}
\tilde{E}_{int}^{Fz}\left[ d\right] =-\frac{\pi R^{3}}{l_{x}^{3}}%
\int_{0}^{\infty }\frac{dq}{q^{2}}e^{-q\frac{d}{R}} \\
\left( 1-e^{-q\frac{W}{R}}\right) ^{2}\left( J_{1}\left( q\right) -\beta
J_{1}\left( q\beta \right) \right) ^{2}.
\end{multline*}%
Thus, the reduced total energy for the Fz state can be expressed as%
\begin{multline*}
\tilde{E}_{tot}^{Fz}=n\tilde{E}_{self}^{Fz}-\frac{\pi R^{3}}{l_{x}^{3}}%
\int_{0}^{\infty }\frac{dq}{q^{2}}e^{-q\frac{W}{R}}\left( 1-e^{-q\frac{W}{R}%
}\right)  \\
\left( J_{1}\left( q\right) -\beta J_{1}\left( q\beta \right) \right)
^{2}g_{z}\left( n,q,\sigma \right) ,
\end{multline*}%
where $g_{z}\left( n,q,\sigma \right) =\frac{\left( n-1\right) e^{q\sigma
}+e^{-\left( n-1\right) q\sigma }-n}{\left( 1-e^{q\sigma }\right) ^{2}}$ and 
$\sigma =\frac{d+W}{R}$.

\subsubsection{Fx state}

For the Fx state, $\mathbf{M}\left( \mathbf{r}\right) $ can be generally
considered as $M_{0}\cos \left[ \left( i-1\right) \theta \right] \hat{x}%
\mathbf{+}M_{0}\sin \left[ \left( i-1\right) \theta \right] \hat{y}$, which
represent a helicoidal magnetic state,\ with $\theta $\ the angle between
the in-plane magnetization of adjacent segments. For the in-plane state, the
exchange and anisotropy contributions to the self energy vanish and the
reduced self energy takes the form \cite{ELA+06}\ 
\begin{multline*}
\tilde{E}_{self}^{Fx}=\frac{\pi R^{3}}{2l_{x}^{3}}\int_{0}^{\infty }\frac{dq%
}{q^{2}}\left( e^{-q\frac{W}{R}}+q\frac{W}{R}-1\right)  \\
\left( J_{1}\left( q\right) -\beta J_{1}\left( q\beta \right) \right) ^{2},
\end{multline*}%
where $J_{1}\left( z\right) $\ is a Bessel function of the first kind. In
order to calculate the interaction energy between the ferromagnetic
segments, we first need to calculate the magnetostatic potential $U\left( 
\mathbf{r}\right) $ of a single tubular structure. The expression for this
potential is given by 
\begin{multline*}
U\left( r,\phi ,z\right) =M_{0}\cos \phi \int_{0}^{\infty }\frac{dk}{k}%
J_{1}\left( kr\right) f\left( k\right)  \\
\left\{ 
\begin{array}{c}
e^{-kz}\sinh \left( k\frac{W}{2}\right) \qquad z>\frac{W}{2} \\ 
\left( 1-e^{-k\frac{W}{2}}\cosh \left( kz\right) \right) \qquad -\frac{W}{2}%
<z<\frac{W}{2} \\ 
e^{kz}\sinh \left( k\frac{W}{2}\right) \qquad z<-\frac{W}{2}%
\end{array}%
\right. ,
\end{multline*}%
where $f\left( k\right) =\left( RJ_{1}\left( kR\right) -aJ_{1}\left(
ka\right) \right) $. Finally, the reduced interaction energy between two
tubular nanostructures is given by%
\begin{multline*}
\tilde{E}_{int}^{Fx}\left[ d,\theta \right] =\frac{\pi R^{3}}{2l_{x}^{3}}%
\cos \theta \int_{0}^{\infty }\frac{dq}{q^{2}}e^{-q\frac{d}{R}} \\
\left( 1-e^{-q\frac{W}{R}}\right) ^{2}\left( J_{1}\left( q\right) -\beta
J_{1}\left( q\beta \right) \right) ^{2}.
\end{multline*}%
Thus, the reduced total energy for the Fx state can be expressed as 
\begin{multline*}
\tilde{E}_{tot}^{Fx}=n\tilde{E}_{self}^{Fx}+\frac{\pi R^{3}}{2l_{x}^{3}}%
\int_{0}^{\infty }\frac{dq}{q^{2}}e^{q\frac{W}{R}}\left( 1-e^{-q\frac{W}{R}%
}\right) ^{2} \\
\left( J_{1}\left( q\right) -\beta J_{1}\left( q\beta \right) \right)
^{2}g_{x}\left( n,q,\sigma ,\theta \right) ,
\end{multline*}%
where $g_{x}\left( n,q,\sigma ,\theta \right)
=\sum_{i=1}^{n-1}\sum_{j=i+1}^{n}e^{-q\sigma \left( j-i\right) }\cos \left[
\left( j-i\right) \theta \right] $ and $\sigma =\frac{d+W}{R}$. From this
expression we can conclude that for zero applied field the total energy of
this state is further reduced for the value $\theta =\pi $, independently of
the value of $n$. Thus, for the particular case when $\theta =\pi $ we
obtain $g_{x}\left( n,q,\sigma ,\pi \right) =-\frac{e^{q\sigma \left(
1-n\right) }\left[ \left( -1\right) ^{n}-e^{nq\sigma }\right] +\left(
1+e^{q\sigma }\right) n}{\left( 1+e^{q\sigma }\right) ^{2}}$.

\subsubsection{Vortex state}

Finally, for the vortex state V, $\mathbf{M}\left( \mathbf{r}\right) $ can
be approximated by $M_{0}\hat{\phi}$, where $\hat{\phi}$ is the azimuthal
unit vector. Due to the condition of perfect flux closure in the vortex
configuration, one magnetic nanostructure in such configuration does not
interact with others, independently of the magnetic configuration of those.
Thus, there is no difference between clockwise and counter-clockwise
directions. Finally, the reduced total energy for the vortex state is given
just by the $n$\ self energies \cite{ELA+06, ELA+07}%
\begin{equation*}
\tilde{E}^{V}=-n\frac{\pi W\ln \beta }{l_{x}}+n\frac{\kappa _{c}}{16}\frac{%
\pi WR^{2}}{l_{x}^{3}}\left( 1-\beta ^{2}\right) .
\end{equation*}%
Here, $\kappa _{c}=2K_{c}/\mu _{0}M_{0}^{2}$.

\subsection{Phase diagram for multisegmented nanorings}

We proceed to investigate the relative stability of the configurations.
Phase diagrams are shown in Fig. 3 for $d=l_{x}$, $\beta =0.5$, and $n=5$. 
\begin{figure}[h]
\begin{center}
\includegraphics[width=8cm]{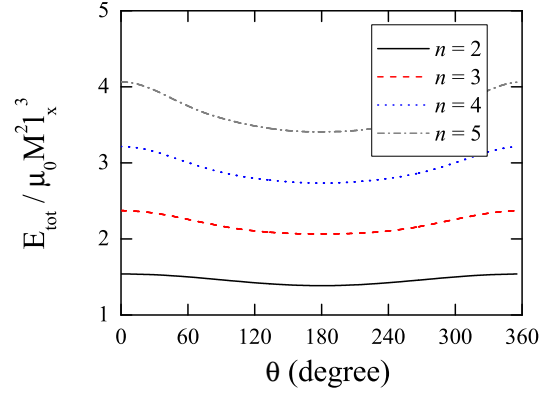}
\end{center}
\caption{Phase diagrams for barcode-type nanostructures giving the regions
in the RW plane where one of the configurations has lower energy. We have
used $\protect\beta =0.5$, $d=l_{x}$, and $n=5$.}
\end{figure}
Anisotropy for four different materials are considered according to values
presented in Table 1. 
\begin{table}[tbph]
\caption{Parameters for different materials taken from Ref. [31]. Uniaxial ($%
K_{u}$) cobalt is denoted with a superscript $^{\ast }$. Iron, permalloy and
nickel have cubic ($K_{c}$) anisotropy.}
\begin{center}
\begin{tabular}{|c|c|c|}
\hline
Material & $K$ (J/m$^{3}$) & $\kappa $ \\ \hline
Iron & $4.8\times 10^{4}$ & $0.0264$ \\ \hline
Cobalt$^{\ast }$ & $4.1\times 10^{5}$ & $0.3329$ \\ \hline
Permalloy & $-3.0\times 10^{2}$ & $-0.0007$ \\ \hline
Nickel & $-4.5\times 10^{3}$ & $-0.0304$ \\ \hline
\end{tabular}%
\end{center}
\end{table}
The diagrams show three regions, corresponding to configurations Fz, Fx, and
V, as in the case of a single nanoring ($n=1$). Notice that for the case of
Co, the existence of a strong uniaxial anisotropy favors the Fz phase,
decreasing the other two phases, specially the V one. In the case of a cubic
anisotropy, the transition lines are similar to the case of a phase diagram
without anisotropy. Because of its very low anisotropy, results for
permalloy describe reasonably well a material with no anisotropy, as it was
pointed in Ref. [27].

Since nanostructures are usually polycrystalline, the crystallographic
orientations of the crystallites are random and, as a consequence, the
average magnetic anisotropy of the particle is very small. In view of that,
it will be neglected in our calculations. \cite{KVL+03, CRE+00}

For different values of $n$ we can determine the ranges of values of the
dimensionless radius $R/l_{x}$ and length $W/l_{x}$ within which one of the
three configurations is of lowest energy. The boundary line between any two
configurations can be obtained by equating the expressions for the
corresponding total energies. Figure 4 illustrates phase diagrams for $%
d=l_{x}$, $\beta =0.5$, and $n=1$ (solid lines), $3$ (dotted lines), and $5$
(dashed lines). It is important to observe that for the Fz and Fx states the
exchange energy is the same. Then, in the absence of applied magnetic fields
and crystalline anisotropies, the dipolar energy is fundamental to obtain
the magnetic configuration of lowest energy. Thus, the dipolar contribution
represents the shape anisotropy that, for multisegmented nanostructures with
a small length (namely nanorings) the low energy state is the quasi uniform
in-plane configuration Fx. \cite{LEA+06} As the length is increased, but
keeping the radius small enough, there is a transition to the out-of-plane
state Fz at a critical length whose value depends on $R$, $\beta $, and the
exchange length $l_{x}$. As the radius is increased, the magnetizations
turns to the vortex configuration at a critical radius depending on the
values of $W$, $\beta $, and $l_{x}$. Finally, by comparing our results we
observe differences in the behavior of the triple point as a function of $n$%
. The triple point occurs for smaller $R/l_{x}$ when $n$ is decreased. 
\begin{figure}[h]
\begin{center}
\includegraphics[width=8cm]{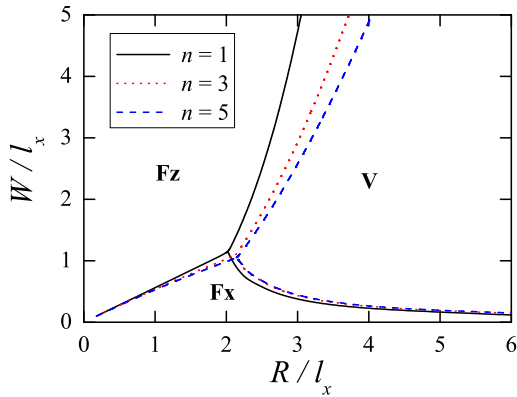}
\end{center}
\caption{Phase diagrams for barcode-type nanostructures with $\protect\beta %
=0.5$ and $d=l_{x}$.}
\end{figure}

Similar to the case of a single ring, the phase diagram changes with $\beta $
\cite{LEA+06}. The dependence of the whole diagram on the value of $n$ can
be investigated by looking at the trajectories of the triple point in the $%
RW $ plane as functions of $\beta $. Such trajectories are shown in Fig. 5
for $d=l_{x}$ and different values of $n$. We remark that the radius $R_{t}$
of the triple point represents the smallest value of $R$ for which the
vortex configurations are stable, and $W_{t}$\ is the biggest value of $W$
for which the in-plane configurations are stable. 
\begin{figure}[h]
\begin{center}
\includegraphics[width=8cm]{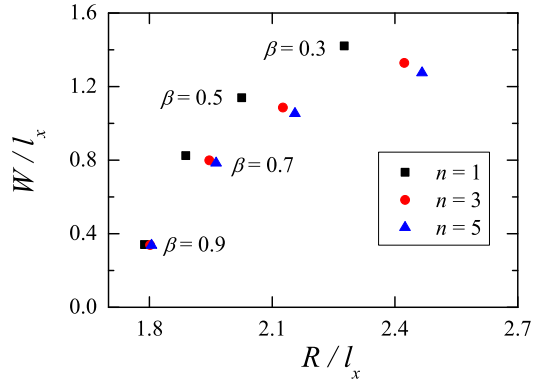}
\end{center}
\caption{Trajectories of the triple point in the phase diagrams in Fig. 4 as
functions of $\protect\beta $, for $n=1$ (squares), 3 (circles), and 5
(triangles).}
\end{figure}

\subsection{Phase diagram for multisegmented nanotubes}

As the multisegmented tubes that motivate this work \cite{LSN+05} satisfy $%
W/R\gg 1$, then the Fx phase can be left out of consideration. Thus, to
obtain an expression for the transition line separating the Fz phase from
the V phase we match the expressions for the energy of these two
configurations. It is important to mention that, for tubes with long radius,
it has been observed a third state which is a mixture of the other two and
has been called \textit{bamboo} or \textit{mixed} state. \cite{Wang05,
LSC+09, LSS+07, CUB+07} As it is known, the consideration of no uniform
magnetic configurations complicates considerably the calculations and for
simplicity, we studied multisegment magnetic nanotubes whose radius are not
big enough to allows the formation of relevant vortex domains at the
extremes of the tube. Figure 6 presents the transition line for $n=1$ and $%
n=2$. To the left of each line Fz state prevails while to the right of the
same line the vortex V configuration is more stable. Labelled dots \textit{%
(continuous)} and \textit{(multisegmented)} in Fig. 6 correspond to the
cases of the two hysteresis curves reported in the experimental paper by Lee 
\textit{et al}. \cite{LSN+05} defined by (\textit{continuous}) $n=1$, $R=150$
nm, $W=16$ $\mu $m, $\beta =0.75$, and $l_{x}=8.225$ nm; (\textit{%
multisegmented}) $n=2$, $R=150$ nm, $W=800$ nm, $d=4800$ nm, $\beta =0.75$,
and $l_{x}=8.225$ nm. It is important to note that the transition line for $%
n=1$ is almost equal to the one with $n=2$. It due to that average distance
between the neighboring Ni segments was big enough ($d=4.8$ $\mu $m)
avoiding thus the interaction between the segments. From this figure we can
conclude that the multisegmented system is well inside the V phase while the
continuous system is inside the Fz phase. It allows us to understand why the
experimental samples show a different magnetic behavior; simply they have
substantial differences in their length of the ferromagnetic segments. 
\begin{figure}[h]
\begin{center}
\includegraphics[width=8cm]{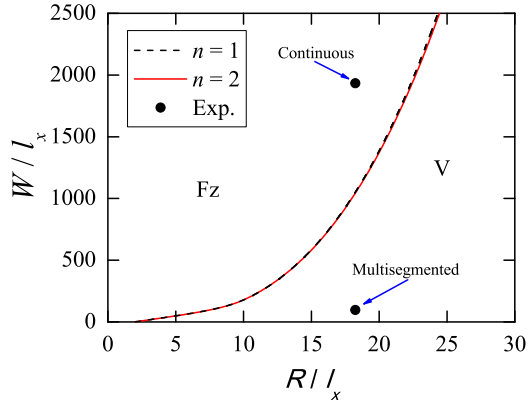}
\end{center}
\caption{Magnetic phase diagrams of non-interacting multisegmented nanotubes
for different values of $n$. The dimensions of the tube, $W$ and $R$, are
normalized to the exchange length $l_{x}$. Experimental points are discussed
in the text. }
\end{figure}

The results presented above may be generalized. We now proceed to
investigate the transition line separating the Fz phase from the V phase. To
obtain an expression for this transition line we match the expressions for
the energy of these two configurations. This leads to $W/l_{x}=\alpha
(\beta) \times R^{3}/l^{3}_{x}$. Function $\alpha (\beta)$ is plotted in
Fig. 7. Care must be applied in the limits of the intervals for $\beta$. In
particular, when $\beta$ goes to 1 we deal with extremely narrow nanotubes,
where eventual surface roughness and thickness irregularities of the
nanotubes become important. On the other side, when $\beta$ goes to zero we
are approaching the limit of a solid cylinder, where the core in the vortex
phase becomes important and must be considered to get the solution. As the
multisegmented nanotubes considered experimentally have $\beta \approx 0.75$%
, we have neglected these two cases.

\begin{figure}[h]
\begin{center}
\includegraphics[width=8cm]{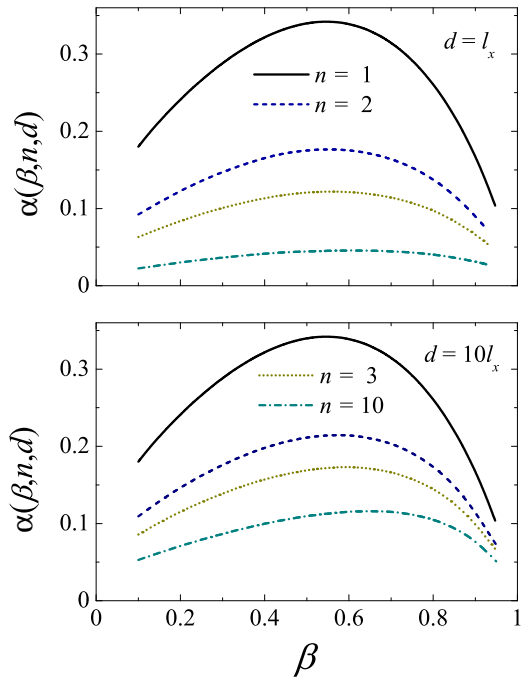}
\end{center}
\caption{Function $\protect\alpha (\protect\beta)$ defining the transition
condition from the phase diagram of a multisegmented nanotube. }
\end{figure}

\section{Conclusions}

In conclusion, we have studied the relative stability of ideal
configurations of magnetic barcode-type tubular nanostructures composed of
alternate ferromagnetic and non-magnetic layers. In such systems we
investigated the size range of the geometric parameters for which different
configurations are of lowest energy. Results are summarized in phase
diagrams which clearly indicate that the magnetic behavior of such
structures can be tailored to meet specific requirements provide a judicious
choice of such parameters is made. The lines separating the magnetic phases
and, in particular, the triple point, are very sensitive to the geometry of
the barcode-type nanostructures. The phase diagrams presented can provide
guidelines for the production of nanostructures with technological purpose.

\section{acknowledgments}

We thank D. Altbir and K. Nielsch for useful discussions. This work was
partially supported by FONDECYT grant numbers 11070010 and 11080246, Financiamiento Basal para Centros Cientificos y Tecnologicos de Excelencia, Millennium Science Initiative under Project P06-022-F, the program
\textquotedblleft Bicentenario en Ciencia y Tecnolog\'{\i}%
a\textquotedblleft\ PBCT under project PSD-031 and the internal Grant
USM-DGIP 11.08.57. We also acknowledge support from the grant program AGCI,
CONICYT, and the program PIIC2009 USM (Chile).

\end{document}